%% file: main.tex
\documentclass[journal]{IEEEtran}
\usepackage[cmex10]{amsmath}
\usepackage{amsthm,amssymb}
\usepackage[utf8]{inputenc}

\usepackage{graphicx}
\usepackage{epstopdf}

\usepackage[labelsep=period]{caption}
\usepackage{subcaption}

\usepackage[ruled,vlined,linesnumbered]{algorithm2e}
\usepackage{multirow}
\usepackage{multicol}
\usepackage{setspace}
\usepackage{tikz}

\usepackage{hyperref}
\usepackage{url}

\usepackage{enumitem}
\setlist[enumerate]{leftmargin=15pt,wide=15pt,}

\usetikzlibrary{arrows}
\allowdisplaybreaks

\input{defs}

\begin{document}

\title{Distributed Video Coding Based on Polar Codes}
\author{
Grigorii Trofimiuk, ~\IEEEmembership{Member,~IEEE,} 
Evgeny Belyaev,  
Peter Trifonov, ~\IEEEmembership{Member,~IEEE}
\thanks{G. Trofimiuk, E. Belyaev and P. Trifonov are with ITMO University, Saint-Petersburg, Russia.
 E-mail: \{gtrofimiuk,eabelyaev,pvtrifonov\}@itmo.ru
 
 This work was supported by the Analytical Center for the Government
of the Russian Federation (IGK 000000D730321P5Q0002), agreement No.
70-2021-00141.}}

\date{\today}
\maketitle

\begin{abstract}
In this letter we  present an improved distributed video coding (DVC) scheme based on polar coding techniques. Firstly,  we adapt log-likelihood ratios (LLRs) for DVC with integer implementation of a discrete cosine transform (DCT). We propose a computationally efficient and numerically stable modification of these LLRs based on the simplified methods of polar codes decoding. We show that on average this approach provides 0.3 dB PSNR gain for DVC with LDPC accumulated (LDPCA) codes. 
Secondly, we introduce the nested shortened polar codes construction algorithm.
We demonstrate that replacement of LDPCA by polar codes improves PSNR by 0.1 dB on average, whereas, for videos with 
relatively 
high motion level, the gain reaches up to 0.23, 0.39 and 0.55 dB for Group of Pictures (GOP) lengths 2, 4 and 8 frames, respectively. Finally, experimental results demonstrate that DVC with polar codes and Tal-Vardy list decoder operates up to two times faster than DVC with LDPCA code and belief propagation (BP) decoder.

%

\end{abstract}


\input{introduction}
\input{background}

\input{polardvc}
\input{numeric}

\input{conclusions}

\bibliographystyle{IEEEtran}
\bibliography{biblio}

\end{document}

%% file: defs.tex
\theoremstyle{plain}
\newtheorem{lemma}{Lemma}

\theoremstyle{remark}

\DeclareMathOperator{\F}{\mathbb F}

\newcommand{\abs}[1]{\left|{#1}\right|}

\newcommand{\ceil}[1]{\left\lceil{#1}\right\rceil}
\renewcommand{\ell}{l}

\newcommand{\floor}[1]{\left\lfloor{#1}\right\rfloor}

\newcommand{\set}[1]{\left\{{#1}\right\}}

\newcommand{\mB}{\mathcal{B}}
\newcommand{\mC}{\mathcal{C}}

\newcommand{\mF}{\mathcal{F}}

\newcommand{\mH}{\mathcal{H}}
\newcommand{\mI}{\mathcal{I}}

\newcommand{\mL}{\mathcal{L}}

\newcommand{\mR}{\mathcal{R}}

\newcommand{\mU}{\mathcal{U}}

\newcommand{\mZ}{\mathcal{Z}}

\newcommand{\bR}{\mathbb{R}}

\newcommand{\bfL}{\mathbf{L}}

\newcommand{\bfP}{\mathbf{P}}
\newcommand{\bfQ}{\mathbf{Q}}



%% file: introduction.tex
\section{Introduction}
DVC is a video compression paradigm driven by emerging applications, such as wireless low-power video
surveillance systems and visual sensor networks. Based on the information-theoretic results of Slepian-Wolf (SW) \cite{slepian1973noiseless} and Wyner-Ziv (WZ) \cite{wyner1976ratedistortion}, it allows  shifting the coding complexity from the encoder to the decoder. Namely, the SW theorem claims that for correlated sources $X$ and $Y$,  a near-lossless compression can be achieved by separate encoding and joint decoding, whereas Wyner and Ziv results 
extends this case to lossy compression, when $Y$ is available at the decoder \cite{ukhanova2010encoder}. In this case, the source 
$Y$ is referred to as \textit{side information} (SI) and 
can be 
considered 
as a noisy version of $X$, obtained via the virtual correlation channel.

Numerous DVC schemes have been proposed in recent years \cite{pereira2009chapter}, including 
transform \cite{artigas2007discover} domain coding with 
LDPCA codes
\cite{varodayan2006rate}, and 
interval overlapped arithmetic coding 
\cite{zhou2019distributed, fang2023qary}. 

Polar codes proposed by Arikan \cite{arikan2009channel} achieve the symmetric capacity of a binary-input memoryless channel. They have  low complexity construction, encoding and decoding algorithms.  Moreover, polar codes were proven to be optimal for lossy source compression and the binary WZ problem \cite{korada2010polar}. Several studies investigated the application of polar codes for the distributed source coding \cite{lv2013novel, yaacoub2016distributed, yaacoub2017systematic}. 

Considering \cite{arikan2009channel}, \cite{korada2010polar}, we propose DVC scheme based on 
with polar codes, which, to the best of the authors' knowledge, is implemented for the first time. The main contributions are the following:
\begin{enumerate}
\item Inspired by the simplified polar code decoding techniques, the modified LLRs for SW decoding and the Laplace model have been proposed. These LLRs are numerically stable, easy to compute and provide on average 0.3 dB PSNR gain for DVC with LDPCA codes and BP decoder.
\item We have introduced the construction of the nested shortened polar codes suitable for DVC. 
In comparison with LDPCA codes, the proposed scheme provides on average 0.1 dB PSNR gain, whereas, for videos with relatively high motion level, the gain reaches up to 0.23, 0.39 and 0.55 dB for GOP lengths 
2, 4 and 8,
frames, 
respectively. For the decoding of polar codes 
we use the successive cancellation list (SCL) decoder, and, as a result, almost twice WZ decoding speed improvement with respect to DVC with LDPCA code and BP decoder is provided.
\end{enumerate}

The rest of the paper is organized as follows. In Section \ref{sBackground} we describe the necessary notations, the basic DVC scheme and the definition of polar codes. Section \ref{sBackground} introduces the proposed approximated LLRs. Section \ref{sNestedPolar} presents nested shortened polar code construction. Section \ref{sExperimental} provides experimental results.  Conclusions
are drawn in Section \ref{sConclusion}.

%% file: background.tex
\section{Background}
\label{sBackground}
\subsection{Notations}
For a positive integer $n$, we denote by $[n]$ the set $\{0,1,\dots ,n-1\}$. 
The vector $u_a^b$ is a subvector $(u_a,u_{a+1},\dots,u_b)$ of a vector $u$. 
For vectors $a$ and $b$, we denote their concatenation by $a.b$.
By $A^{\otimes m}$ we denote the $m$-fold Kronecker product of the matrix $A$ by itself. 
\subsection{Basic Distributed Video Coding Framework}
\label{ssBasicDVC}
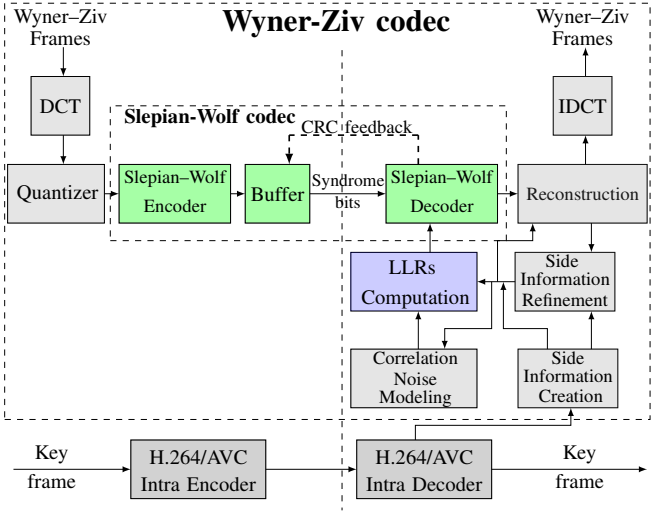
\begin{figure}[ht]
\centering
\scalebox{0.78}{
\input{TikzWZ2}
}
\caption{Basic Distributed Video Coding scheme}
\label{fCodecDiagram}
\end{figure}

The DVC architecture used in this work 
is depicted in Fig. \ref{fCodecDiagram} and operates as follows \cite{chiu2012hybrid}.  An input video sequence is split into GOP, where the first frame is Intra encoded (\textit{key frame}) by H.264/AVC (without motion prediction) and the remaining frames are WZ encoded 
(\textit{WZ frame}).  The part of DVC responsible for lossy compression of a WZ frame is designated as \textit{WZ codec} consisting of the following steps:
\subsubsection{Discrete cosine transform}
A $w \times h$ WZ frame is divided into non-overlapped $4 \times 4$ blocks, which the integer DCT is applied to. By collecting element $(i, j)$ in each block, the DCT band $X^{(\phi)}$ with length $n = w \cdot h /16$ is obtained, where $\phi = s(i,j)$, and $s(i,j)\in [16]$ is the order of DCT block scanning.

\subsubsection{Quantization}
\label{sssQuant}
Band $X^{(\phi)}$ is quantized into a vector 
$\widehat X^{(\phi)}_i = Q_{\phi}(X^{(\phi)}_i), i \in [n]$.
 We consider an $M_{\phi}$-level quantizer $Q_\phi(x)$, 
which maps a number $x$ to a binary label $b_0^{\mu_\phi-1}$ of $\mu_\phi$ bits, $M_\phi \leq 2^{\mu_\phi}$. WZ codec uses eight different $4\times 4$ quantization matrices $\bfQ_f$ \cite{pereira2009chapter}, where 
$\mu_{s(i,j)} = \bfQ_f[i,j]$. The index $f$ determines the quality of the decoded WZ frame, resulting in different rate-distortion (RD) points. In our scheme, we use quantizers from \cite{pereira2009chapter}, but with fixed quantization levels.
\subsubsection{Slepian-Wolf Encoding}
\label{sssSWE}
After quantization, bitplanes $b^{(l,\phi)}, l \in [\mu_\phi],$ are collected, where $b^{(l,\phi)}_i$ is given by the $l$-th bit of $\widehat X^{(\phi)}_i$. Let us consider a chain of $\omega$ nested $(n, k_i)$ linear codes $\mC_i$,  
$\mC_{\omega-1} \subset \mC_{w-2} \subset \cdots \subset \mC_{0}$, $k_{w-1} = 0$, $k_{-1} = n$, with nested parity-check matrices $\mH_i,$ i.e.
$
\mH_{i+1} = \left( \begin{array}{c}
 \mH_i  \\
 \hline
 D_i 
\end{array}
\right),
$
where 
$D_i$ is $(k_i-k_{i+1}) \times n$ matrix,
Then, $n$-length syndromes $h^{(b,l)} = b^{(l,\phi)} \mH_{\omega-1}^{\top}$ are computed and stored in a buffer. After that, for each 
$b^{(l,\phi)}$, the $t$-bit CRC is computed and sent to the decoder together with $h_0^{k_0}$.
\subsubsection{Side Information Creation}
\label{sssBackSI}

On the WZ decoder side, the key frames are decoded and the SI frame $Y$ is created by interpolating the closest  
frames already decoded \cite{ascenso2005improving}. Then, DCT $S$ of $Y$ is obtained. WZ problem implies \cite{cheng2005successive} $X^{(\phi)} = S^{(\phi)} + V^{(\phi)}$, where $V^{(\phi)}$ is 
a correlation noise (CN) sample. A Laplace distribution with  mean $0$ and 
variance $2/\alpha_\phi^2$ is commonly used for CN modeling, which parameters are estimated online \cite{brites2008correlation}.
\subsubsection{Slepian-Wolf Decoding}
\label{sssSWDec}
 The decoder of nested codes $\mC_i$ is used to successively obtain bitplanes $b^{(l,\phi)}$. Starting from the bitplane $b^{(\mu_\phi-1,\phi)}$, for each $b^{(l,\phi)}$ at level $l$, it calculates the input LLRs $L_0^{n-1}$ regarding $S^{(\phi)}$, $\alpha_\phi$ and the already decoded bitplanes \cite{cheng2005successive}. 
Given these LLRs, starting from $i = 0$, the decoder tries to obtain such $v_0^{n-1}$ as $h_0^{n-k_i-1} = v_0^{n-1} \mH_{i}^{\top}$. If the CRC of $v_0^{n-1}$ does not match the CRC of the corresponding bitplane, then the additional chunk $h_{n-k_{i-1}}^{n-k_i-1}$ of the syndrome is requested from the WZ encoder buffer via a feedback channel. If decoding of $\mC_{w-2}$ fails, the decoder receives the last chunk, and the bitplane $b^{(l,\phi)}$ is recovered as $h_0^{n-1} (\mH_{\omega-1}^\top)^{-1}$.    

The algorithm described above can be viewed as  multistage decoding of a multilevel code \cite{wachsmann99multilevel} with $\mu_\phi$ levels and bit reversed natural labeling. 
The channel codec used for lossless compression of $\widehat X^{(\phi)}$ is usually denoted as the Slepian-Wolf codec (SWC). 
The scheme \cite{chiu2012hybrid} uses the LDPCA code with the sum-product BP decoder. The decoder also uses the received syndrome $h_0^{n-k_i-1}$ for early termination when it converges five times in a row to the same wrong hard decisions $\widehat c$, i.e. 
$\widehat c \mH_{i}^{\top} \ne h_0^{n-k_i-1}.$ 
\subsubsection{Reconstruction}
The decoded bitplanes form a quantized band $\widehat X^{(\phi)}$. The reconstruction of $X^{(\phi)}$ is performed considering both SI $Y$ and $\widehat X^{(\phi)}$ \cite{kubasov2007optimal}. After $X^{(\phi)}$ is reconstructed, SI $Y$ is refined \cite{martins2009refining} and used to update the CN parameter $\alpha$.

\subsection{Polar Codes}
\label{ssPolarCodes}
A $(N=2^m,k)$ polar code \cite{arikan2009channel} over $\F_2$ is a set of vectors $c_0^{N-1}=u_0^{N-1}F_{m}$,  where 
$
F_m = \left(
\arraycolsep=1.15pt\def\arraystretch{0.5}
\begin{array}{cc}
1&0\\
1&1
\end{array} \right)^{\otimes m}
$
is a matrix of the polarizing transformation, $u_i, i\in \mathcal F \subset[N],$ are set to some predefined values (frozen set), e.g. zero,  $|\mF| = N - k$, and the remaining values $u_i$ are set to the payload data. 
It can be shown that the matrix $F_m$ together with a memoryless output symmetric channel $W(y|x)$ gives rise to synthetic bit subchannels $W_m^{(i)} = W_m^{(i)}(y_0^{n-1},u_0^{i-1}|u_i)$
with transition probability functions 
$$W_m^{(i)}(y_0^{n-1},u_0^{i-1}|u_i)=\frac{1}{2^{N-1}}
\sum_{u_{i+1}^{N-1}}\prod_{j=0}^{n-1}W(y_j|(u_0^{N-1}F_{m})_j).$$
Classical polar codes are obtained by taking $\mF$ as the set of $N-k$ indices $i$ of bit subchannels $W_m^{(i)}$ with the highest error probability or the Bhattacharyya parameter $Z(W_m^{(i)})$ \cite{arikan2009channel}.

%% file: TikzWZ2.tex
\begin{tikzpicture}

\draw[fill=gray,opacity=0.2]  (-3.45,4.5) rectangle (-1.8,3.5);
\draw  (-3.45,4.5) rectangle (-1.8,3.5);
\node at (-2.6,4) {Quantizer};

\draw[-latex] (1.7,4) -- (3,4);

\begin{scope}
\draw[fill=green,opacity=0.35]  (-1.55,4.5) rectangle (0.35,3.5);
\draw  (-1.55,4.5) rectangle (0.35,3.5);
\node[scale = 0.9] at (-0.6,4.25) {Slepian–Wolf};
\node[scale = 0.9] at (-0.6,3.8) {Encoder};
\end{scope}

\begin{scope}[shift={(4.1,0)}]
\draw[fill=green,opacity=0.35]  (-1.1,4.5) rectangle (0.8,3.5);
\draw  (-1.1,4.5) rectangle (0.8,3.5);
\node[scale = 0.9] at (-0.15,4.25) {Slepian–Wolf};
\node[scale = 0.9] at (-0.15,3.8) {Decoder};
\end{scope}
\node[scale = 0.85] at (2.35,4.2) {Syndrome};
\node[scale = 0.85] at (2.35,3.85) {bits};
\draw[fill=green,opacity=0.35]  (0.6,4.5) rectangle (1.7,3.5);
\draw  (0.6,4.5) rectangle (1.7,3.5);
\node at (1.15,4) {Buffer};

\draw[-latex] (0.35,4) -- (0.6,4);
\node[scale = 0.9] at (2.5,5.15) {CRC feedback};
\draw[dashed, thick]  (3.55,5) -- (1.35,5);
\draw[dashed, thick] (3.55,4.5) -- (3.55,5);
\draw[dashed, -latex, thick] (1.35,5) -- (1.35,4.5);

\node[scale = 0.9] at (6.35,4) {Reconstruction};
\draw  (5.25,4.5) rectangle (7.45,3.5);
\draw[fill=gray,opacity=0.2]  (5.25,4.5) rectangle (7.45,3.5);
\draw[-latex] (4.9,4) -- (5.25,4);

\begin{scope}[shift={(3.4,0)}]
\node at (-5.95,7) {Wyner–Ziv};
\node at (-5.95,6.65) {Frames};
\draw[-latex] (-5.9,6.5) -- (-5.9,6);
\draw[-latex] (-5.9,5) -- (-5.9,4.5);
\draw[fill=gray,opacity=0.2]  (-6.45,6) rectangle (-5.45,5);
\draw  (-6.45,6) rectangle (-5.45,5);
\node at (-5.95,5.5) {DCT};
\end{scope}

\begin{scope}[shift={(10.6,0)}]
\node at (-4.25,7) {Wyner–Ziv};
\node at (-4.25,6.65) {Frames};
\draw[-latex] (-4.2,6) --  (-4.2,6.5);
\draw[-latex] (-4.2,4.5) -- (-4.2,5);
\draw  (-4.75,6) rectangle (-3.75,5);
\draw[fill=gray,opacity=0.2] (-4.75,6) rectangle (-3.75,5);
\node at (-4.25,5.5) {IDCT};
\end{scope}

\begin{scope}[shift={(0,-1.05)}]
\begin{scope}[shift={(2.25,-1.15)}]
\draw[fill=gray,opacity=0.35]  (-3.6,2) rectangle (-1.3,1);
\draw  (-3.6,2) rectangle (-1.3,1);
\node at (-2.45,1.7) {H.264/AVC};
\node at (-2.45,1.25) {Intra Encoder};
\end{scope}

\node at (-2.7,0.6) {Key};
\node at (-2.7,0.15) {frame};
\draw[-latex] (-3.35,0.35) -- (-1.35,0.35);

\begin{scope}[shift={(7.15,-1.15)}]
\draw[fill=gray,opacity=0.35]  (-4.65,2) rectangle (-2.35,1);
\draw  (-4.65,2) rectangle (-2.35,1);
\node at (-3.5,1.7) {H.264/AVC};
\node at (-3.5,1.25) {Intra Decoder};
\end{scope}
\draw[-latex] (0.95,0.35) -- (2.5,0.35);
\draw[-latex] (4.8,0.35) -- (7.45,0.35);

\node at (6.3,0.6) {Key};
\node at (6.3,0.15) {frame};
\end{scope}

\draw  (2.4,3) rectangle (4.55,2);
\draw[fill=blue,opacity=0.2]  (2.4,3) rectangle (4.55,2);
\node at (3.45,2.75) {LLRs};
\node at (3.5,2.2) {Computation};

\draw[dashed] (2.25,6.45) -- (2.25,-1.4);
\draw[-latex] (3.75,3) -- (3.75,3.5);

\draw  (5.25,1.35) rectangle (6.95,0.35);
\draw[fill=gray,opacity=0.2]  (5.25,1.35) rectangle (6.95,0.35);

\node[scale = 0.9] at (6.1,1.2) {Side};
\node[scale = 0.9] at (6.1,0.9) {Information};
\node[scale = 0.9] at (6.15,0.55) {Creation};

\draw[-latex] (3.55,1.35) -- (3.55,2);

\begin{scope}[shift={(-2.5,-0.15)}]

\draw[fill=gray,opacity=0.2]  (4.9,1.5) rectangle (7.05,0.5);
\draw  (4.9,1.5) rectangle (7.05,0.5);

\node[scale = 0.9] at (6,1.35) {Correlation};
\node[scale = 0.9] at (6,1) {Noise};
\node[scale = 0.9] at (6,0.65) {Modeling};
\end{scope}

\draw  (5.2,3) rectangle (6.9,2);
\draw[fill=gray,opacity=0.2]  (5.2,3) rectangle (6.9,2);

\node[scale = 0.9] at (6,2.85) {Side};
\node[scale = 0.9] at (6.05,2.55) {Information};
\node[scale = 0.9] at (6.05,2.2) {Refinement};

\draw[-latex] (6.5,3.5) -- (6.5,3);

\draw[-latex] (6.5,1.35) -- (6.5,2) ;

\draw[-latex] (5.2,2.5) -- (4.55,2.5);
\draw[-latex] (5.75,1.35) -- (5.75,1.7) -- (5,1.7) -- (5,2.5);
\draw[-latex] (4.8,2.5) -- (4.8,1.7) -- (4,1.7) -- (4,1.35);
\draw[-latex] (3.5,-0.2) -- (3.5,0.05) -- (6.15,0.05) -- (6.15,0.35);
\draw[-latex] (-1.8,4) -- (-1.55,4);
\draw[-latex] (4.9,2.5) -- (4.9,3.2) -- (5.5,3.2) -- (5.5,3.5);

\draw[dashed]  (-3.5,7.25) rectangle (7.5,0.15);
\node[scale = 1.5] at (2.15,6.9) {\textbf{Wyner-Ziv codec}};
\draw[dashed]  (-1.7,5.5) rectangle (5.05,3.2);
\node at (0,5.3) {\textbf{Slepian-Wolf codec}};
\end{tikzpicture}

%% file: polardvc.tex
\section{ Log-likelihood Ratios for Integer DCT}
\label{sApproximatedLLRs}
\subsection{Log-likelihood ratios in basic DVC scheme}
Let $s_0^{n-1}$ be a DCT band of SI $S$. We assume
that the quantizer $Q$ outputs such labels $b_0^{\mu-1}$ that multistage decoding starts from bits $b_0$ and ends at $b_{\mu-1}$. Let us define a set 
$$\mB(b_0^{l-1},\mU) = \set{x| x\in \mU, \bar b = Q(x), \bar b_0^{l-1} = b_0^{l-1}}.$$ For the Wyner-Ziv problem (see Section \ref{sssBackSI}), floating point DCT implementation, and Laplace model \cite{kubasov2007optimal}, we have \cite{cheng2005successive}

\begin{align}
\bfP(b_0^{l}|s_i) = \!\!\!
\int\limits_{\mB(b_0^{l-1}, \bR)} \!\!\!P(x|s_i)dx = 
\int\limits_{\mB(b_0^{l-1},\bR)}\frac{\alpha}{2}e^{-\alpha\abs{x-s_i}} dx.
\end{align}
Probabilities $\bfP(b_0^{l}|s_i)$ can be analytically computed and used to obtain soft-input LLRs of the SW decoder
\begin{equation}
\bfL(\widehat b_0^{l-1}|s_i) = 
\log \frac{\bfP(\widehat b_0^{l-1}.0|s_i)}
{\bfP(\widehat b_0^{l-1}.1|s_i)},
\label{fOptimalLLRs}
\end{equation}
where $\widehat b_0^{l-1}$ are known bits of the $i$-th symbol. 

We denote LLRs \eqref{fOptimalLLRs} as the \textit{basic} ones, since they are implemented in \cite{chiu2012hybrid}.
\subsection{Proposed log-likelihood ratios}
\label{ssProposedLLRs}
Recall that the DVC scheme, considered in this letter, uses an integer implementation of DCT. Suppose that $X^{(\phi)}_i \in \mI$. For this setting, we propose to define the following probability
\begin{equation}
\label{fSumProb}
P(b_0^{l}|s_i) = \Delta^{-1}
\sum_{x \in {\mB(b_0^{l-1},\mI)}} P(x|s_i),
\end{equation}
where $\Delta = \sum_{x\in \mI} P(x|s_i)$ is the normalization coefficient.
As it was observed in \cite{miloslavskaya2014sequential, trifonov2022multilevel}, in the context of polar and multilevel codes, the decoding can be simplified by replacing summation with maximization in the expression for bit subchannel probabilities. Therefore, we introduce the following values:
\begin{equation}
\label{fMaxProb}
\widetilde P(b_0^{l}|s_i) = \Delta^{-1}
\max_{x \in {\mB(b_0^{l-1},\mI)}} P(x|s_i).
\end{equation}
Then, the LLRs of the values \eqref{fMaxProb} and the Laplace model are given by 
\begin{align}
\label{fApproxLLRs}
&L(b_0^{l-1}|s_i) = \log \frac{\widetilde P(b_0^{l-1}.0|s_i)}
{\widetilde P( b_1^{l-1}.1|s_i)} = 
\log \frac{
\max\limits_{x \in {\mB(b_0^{l-1}.0,\mI)}} P(x|s_i)
}
{
\max\limits_{x \in {\mB(b_0^{l-1}.1,\mI)}} P(x|s_i)
} \nonumber\\
&\quad = \alpha(R^{(1)}(b_0^{l-1}|s_i) - R^{(0)}(b_0^{l-1}|s_i)) = \alpha R(b_0^{l-1}|s_i).
\end{align}
where 
$$R^{(j)}(b_0^{l-1}|s_i) = \min \limits_{x \in \mB(b_0^{l-1}.j,\mI)}|x-s_i|.$$ Let us observe that $R(b_0^{l-1}|s_i)$ is a piecewise linear function that admits  simple and numerically stable computation. 
 Namely, for the $(0, 0)$ DCT band, $2^\mu$-level uniform scalar quantizer is used \cite{pereira2009chapter}, which quantizes an integer 
$A_{j-1} \leq x < A_j$ into a label $\mL^{(j)}$, where $A_j = (j+1)  2^{\beta - \mu}$, $j \in [2^{\mu}]$, $A_{-1} = -\infty$, $A_\mu = \infty$, and $\sum_{k=0}^{\mu-1} \mL^{(j)}_k 2^{\mu-k-1} = j$. For the vector $b_0^{l-1}$, we define the values $\gamma = \beta - l - 1$ and $a =  2^{\gamma} \cdot \sum_{j=0}^{l-1} b_j \cdot 2^{(l-j)}$. It can be verified that 
\begin{equation}
\label{eRLLR}
R(b_0^{l-1}|s_i) = 
\begin{cases}
2^{\gamma}, s_i < a,\\
-2^{\gamma}, s_i \geq a+ 2^{\gamma+1},\\
2^{\gamma} \!\!-\!(s_i\!-\!a)\!-\!\floor{(s_i\!-\!a)\!/2^{\gamma}\!}\!,\text{otherwise.}\\
\end{cases}
\end{equation}
One can see that computation of \eqref{eRLLR} requires only integer comparisons and summations. For other DCT bands, which use uniform scalar quantizers with doubled zero interval \cite{pereira2009chapter},  similar expressions for $R(b_0^{l-1}|s_i)$ can be also obtained.
\section{Construction of rate-compatible polar codes}
\label{sNestedPolar}
As described in Section \ref{sssSWE}, 
SWC requires the construction of a chain of nested linear codes.  Polar code construction allows length $2^t$ only. To obtain a polar code of arbitrary length $n$, we consider the shortening of $2^t$ length polar code at the last $2^t-n$ positions, $t = \ceil{\log_2n}$.  Then, to construct nested codes, we propose to compute a sequence $\mR$ of integers  such that the bit subchannel $W_m^{(\mR_j)}$ becomes sufficiently unreliable after the subchannel $W_m^{(\mR_{j+1})}$, while degrading the quality of the underlying channel. As a result, the frozen set for the $(n, k)$ code in the chain is given by $\set{\mR_0^{n-k-1}}$.

Consider a family of binary discrete memoryless channels (B-DMCs) $W(\sigma)$ indexed by a parameter $\sigma \geq 0$ and ordered with respect to degradation, i.e. $W(\sigma_1) \preceq W(\sigma_2) \Leftrightarrow \sigma_1 > \sigma_2$.
 
\begin{lemma}[\cite{korada2010polar}]
\label{lBD}
Let $W:\{0,1\} \rightarrow \mathcal{Y}$ and $W^{\prime}:\{0,1\} \rightarrow \mathcal{Y}^{\prime}$ be two B-DMCs such that $W \preceq W^{\prime}$, then for all $i, W_N^{(i)} \preceq W_N^{\prime(i)}$ and hence $Z(W_m^{(i)}) \geq Z(W_m^{\prime(i)})$.
\end{lemma}
By $\mZ_i^{(\sigma)}$ we denote a Bhattacharyya parameter of the $i$-th bit subchannel after the application of polarizing transform to $W(\sigma)$. 
The DVC scheme implies that the variance of the correlation channel (i.e. quality of SI) differs from frame to frame; thus, nested codes for SW coding should be suitable for different channel conditions.
That is, relying on Lemma \ref{lBD}, we propose to construct nested polar codes by varying channel conditions. 

\begin{algorithm}[ht]
\SetNoFillComment
\caption{\texttt{GetReliabilitySequence}$(n, T, \epsilon)$}
\label{alg_RelSeq}
$\mR \gets \bf{0}^{n}$; \tcp{Reliability sequence vector}
$\mF \gets \emptyset$; \tcp{Indices already in $\mR$} 
\For{$i \in [n]$}{
Find such $\sigma$ that $T-\epsilon\leq 
\min_{j \in [n]\setminus \mF} \mZ_j^{(\sigma)}
 \leq T+\epsilon$;\label{lZOpt}\\
$j^{\star} \gets \arg \min_{j \in [n]\setminus \mF} \mZ_j^{(\sigma)}$;\\
$\mR_{n-i-1} \gets j^{\star}, \mF \gets \mF \cup \set{j^{\star}}$;\\
}
\Return $\mR$
\end{algorithm}
Alg. \ref{alg_RelSeq} presents the proposed nested code construction. It changes $\sigma$ to add the next index to the reliability 
sequence $\mR_0^{n-1}$. At line \ref{lZOpt} we adjust the channel quality parameter $\sigma$ so that the minimum value of the Bhattacharyya parameter among the subchannels, which were not added to $\mR$, belongs to the interval $[T-\epsilon; T+\epsilon],T \in (0;1), 0 \leq \epsilon < T$. Lemma \ref{lBD} ensures that $\mZ_{\mR_{i+1}}^{(\sigma)} < \mZ_{\mR_{i}}^{(\sigma)}$ for any fixed 
$\sigma > 0$ and $i \in [n]$.

In this letter we set $W(\sigma)$ as AWGN channel with variance $\sigma^2$. The values of $\mZ_i^{(\sigma)}$ for shortened polar codes can be estimated by the Tal-Vardy method \cite{tal2013how}, density evolution \cite{mori2009performance} or the Gaussian approximation (GA) \cite{trifonov2012efficient}, which is here used  due to its computational efficiency.

It should be noted that, for the shortened polar code (see Section \ref{ssPolarCodes}), $(\mH_{w-1}^{\top})^{-1}$ is a submatrix of the Arikan matrix $F_t$, $t = \ceil{(\log_2 n)}$. Due to
its recursive structure, the original bitplane $b^{(\phi,l)} = h_0^{n-1} (\mH_{w-1}^{\top})^{-1}$  can be obtained in $O(2^t \cdot t)$ operations. This is much simpler in comparison with the case of LDPCA codes, which, to the best of authors' knowledge, have no fast algorithms for generator matrix multiplication.



In this paper we consider the usage of a single chain of nested codes for all DCT bands and bitplanes. The joint optimization of quantizers and SWC lies out of the scope of this paper and is a topic of future research.


%% file: numeric.tex
\section{Experimental Results}
\label{sExperimental}
Experimental results were obtained for the luma component (grayscale) of 27 test videos\footnote{Akiyo, Bowing, Bridge-close,
Bridge-far, Carphone, City, Coast Guard, Container,
Crew, Deadline, Flower, Football, Foreman, Hall Monitor,
Harbour, Ice, Mobile, Mother-daughter, News, 
Pamphlet, Paris, Sign Irene, Silent, Soccer, Students,
Tempete and Waterfall.} from~\cite{Xiph} with frame resolution 
$176\times 144$ and frame rate 15~Hz. This implies SWC of length $n = 1584$. We used regular degree-3 LDPCA code (typical for DVC) \cite{varodayan2006rate, ldpcaref}, decoded by BP with maximum 100 iterations together with 12 bit CRC for feedback. For both LDPCA and shortened polar codes we used a chain of 66 nested codes of dimensions $1536, 1512, \dots, 48, 24,0$. Shortened polar codes were decoded by the SCL decoder \cite{tal2015list} with list size $L = 32$ and 28 bit CRC. On the one hand, in contrast to the BP decoder, the SCL decoder always produces a list of valid codewords, and we need more CRC bits for error detection. On the other hand, we check the CRC for each codeword in this list, which reduces overall bitrate. We constructed polar codes by the proposed Alg. \ref{alg_RelSeq} with $T = 10^{-3}, \epsilon = 10^{-4}$, which were chosen by examining RD performance for different $T$.



\begin{figure}[ht]
\centering
\includegraphics[width=0.95\linewidth]{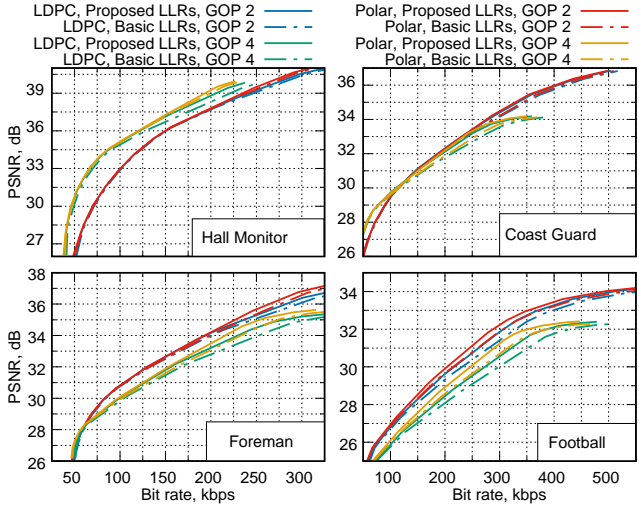}
\caption{RD comparison of different SWCs}
\label{fRD}
\end{figure}

We obtained different RD points by changing the index $f$ (\textit{QP WZ}) of a quantization matrix $\bfQ_f$ (see Section \ref{sssQuant}) and the quantification parameter
(\textit{QP Intra}) of H264/AVC. 
Fig.~\ref{fRD} demonstrates the RD performance of DVC with the considered codes for four test video sequences with different motion levels. It can be seen that for video sequences with low motion levels (for example, see Hall Monitor) all the considered codes provide similar performance.
However, with the increase of GOP length and motion level, the similarity between key frames decreases. As a result, the quality of SI is degraded, leading to a higher bit error rate in the virtual channel. In this case, the SWC with the proposed LLRs \eqref{fApproxLLRs} provides better error correction ability, i.e., it requires smaller syndrome bit length for a given video quality level. Herewith, the proposed polar code outperforms the LDPCA one. 
\begin{figure}[ht]
\centering
\includegraphics[width=0.95\linewidth]{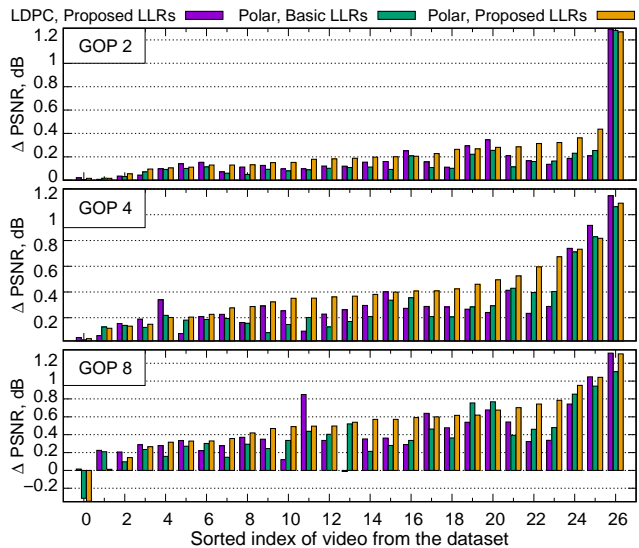}
\caption{PSNR gain related to the basic DVC scheme}
\label{fBJ}
\end{figure}

Fig.~\ref{fBJ} shows the
quality improvement achieved by the proposed DVC scheme relative to the basic one (LDPCA code with basic LLRs) according to the 
Bjontegaard $\Delta$PSNR metric~\cite{Bjontegaard2001}. Videos are arranged in the increasing order of $\Delta$PSNR for polar codes and proposed LLRs. It can be observed that the proposed modified LLRs (see Section \ref{ssProposedLLRs}) improve RD performance for both LDPCA and polar codes by 0.3 dB on average, while the proposed polar codes provide additional 0.1 dB gain on average and, for videos with relatively high motion level, such as Football, up to 0.23, 0.39 and 0.55 dB for GOP  2, 4 and 8 frames, respectively.


\begin{table}[ht]
\centering
\caption{WZ decoding time $\tau$ with different SWC, GOP 4}
\label{tDecTime}
\resizebox{0.99\columnwidth}{!}{%
\begin{tabular}{|c||ccc||ccc||ccc||ccc|}
\hline
Video                                                            & \multicolumn{3}{c||}{football}                                                                                                 & \multicolumn{3}{c||}{foreman}                                                                                                  & \multicolumn{3}{c||}{coastguard}                                                                                               & \multicolumn{3}{c|}{hall\_monitor}                                                                                            \\ \hline
\multirow{2}{*}{\begin{tabular}[c]{@{}c@{}}QP\\ WZ\end{tabular}} & \multicolumn{1}{c|}{\multirow{2}{*}{\begin{tabular}[c]{@{}c@{}}QP\\ Intra\end{tabular}}} & \multicolumn{2}{c||}{$\tau$, sec}   & \multicolumn{1}{c|}{\multirow{2}{*}{\begin{tabular}[c]{@{}c@{}}QP\\ Intra\end{tabular}}} & \multicolumn{2}{c||}{$\tau$, sec}   & \multicolumn{1}{c|}{\multirow{2}{*}{\begin{tabular}[c]{@{}c@{}}QP\\ Intra\end{tabular}}} & \multicolumn{2}{c||}{$\tau$, sec}   & \multicolumn{1}{c|}{\multirow{2}{*}{\begin{tabular}[c]{@{}c@{}}QP\\ Intra\end{tabular}}} & \multicolumn{2}{c|}{$\tau$, sec}   \\ \cline{3-4} \cline{6-7} \cline{9-10} \cline{12-13} 
                                                                 & \multicolumn{1}{c|}{}                                                                    & \multicolumn{1}{c|}{LDPC}  & Polar & \multicolumn{1}{c|}{}                                                                    & \multicolumn{1}{c|}{LDPC}  & Polar & \multicolumn{1}{c|}{}                                                                    & \multicolumn{1}{c|}{LDPC}  & Polar & \multicolumn{1}{c|}{}                                                                    & \multicolumn{1}{c|}{LDPC}  & Polar \\ \hline
0                                                                & \multicolumn{1}{c|}{41}                                                                  & \multicolumn{1}{c|}{168.3} & 75.1  & \multicolumn{1}{c|}{41}                                                                  & \multicolumn{1}{c|}{110.2} & 56.7  & \multicolumn{1}{c|}{39}                                                                  & \multicolumn{1}{c|}{53.7}  & 35.2  & \multicolumn{1}{c|}{35}                                                                  & \multicolumn{1}{c|}{27.3}  & 26.0  \\ \hline
3                                                                & \multicolumn{1}{c|}{39}                                                                  & \multicolumn{1}{c|}{261.4} & 120.3 & \multicolumn{1}{c|}{35}                                                                  & \multicolumn{1}{c|}{158.8} & 84.2  & \multicolumn{1}{c|}{34}                                                                  & \multicolumn{1}{c|}{79.4}  & 57.5  & \multicolumn{1}{c|}{29}                                                                  & \multicolumn{1}{c|}{39.8}  & 37.5  \\ \hline
6                                                                & \multicolumn{1}{c|}{33}                                                                  & \multicolumn{1}{c|}{412.5} & 194.6 & \multicolumn{1}{c|}{31}                                                                  & \multicolumn{1}{c|}{240.3} & 133.8 & \multicolumn{1}{c|}{29}                                                                  & \multicolumn{1}{c|}{138.4} & 99.2  & \multicolumn{1}{c|}{26}                                                                  & \multicolumn{1}{c|}{66.5}  & 66.2  \\ \hline
7                                                                & \multicolumn{1}{c|}{26}                                                                  & \multicolumn{1}{c|}{697.9} & 319.9 & \multicolumn{1}{c|}{26}                                                                  & \multicolumn{1}{c|}{408.4} & 221.7 & \multicolumn{1}{c|}{23}                                                                  & \multicolumn{1}{c|}{283.3} & 187.7 & \multicolumn{1}{c|}{23}                                                                  & \multicolumn{1}{c|}{132.3} & 108.4 \\ \hline
\end{tabular}%
}
\end{table}

Table \ref{tDecTime} compares the WZ decoding time, labeled $\tau$, of DVC with different SWC, measured on the CPU Intel Core i7-9700K. We can see that the DVC decoder based on polar codes is almost twice faster than with LDPCA codes.


%% file: conclusions.tex
\section{Conclusions}
\label{sConclusion}
In this letter a distributed video coding scheme with polar codes is proposed, which employs nested shortened polar codes as well as modified log-likelihood ratios for the multistage decoder and Laplace model. The proposed scheme provides both rate-distortion and decoding speed improvement. The highest PSNR gain is achieved for videos with high motion level.

%% file: main.bbl
\begin{thebibliography}{10}
\providecommand{\url}[1]{#1}
\csname url@samestyle\endcsname
\providecommand{\newblock}{\relax}
\providecommand{\bibinfo}[2]{#2}
\providecommand{\BIBentrySTDinterwordspacing}{\spaceskip=0pt\relax}
\providecommand{\BIBentryALTinterwordstretchfactor}{4}
\providecommand{\BIBentryALTinterwordspacing}{\spaceskip=\fontdimen2\font plus
\BIBentryALTinterwordstretchfactor\fontdimen3\font minus
  \fontdimen4\font\relax}
\providecommand{\BIBforeignlanguage}[2]{{%
\expandafter\ifx\csname l@#1\endcsname\relax
\typeout{** WARNING: IEEEtran.bst: No hyphenation pattern has been}%
\typeout{** loaded for the language `#1'. Using the pattern for}%
\typeout{** the default language instead.}%
\else
\language=\csname l@#1\endcsname
\fi
#2}}
\providecommand{\BIBdecl}{\relax}
\BIBdecl

\bibitem{slepian1973noiseless}
D.~Slepian and J.~Wolf, ``Noiseless coding of correlated information sources,''
  \emph{IEEE Trans. Inf. Theory}, vol.~19, no.~4, pp. 471--480, July 1973.

\bibitem{wyner1976ratedistortion}
A.~Wyner and J.~Ziv, ``The rate-distortion function for source coding with side
  information at the decoder,'' \emph{IEEE Trans. Inf. Theory}, vol.~22, no.~1,
  pp. 1--10, Jan. 1976.

\bibitem{ukhanova2010encoder}
A.~Ukhanova, E.~Belyaev, and S.~Forchhammer, ``Encoder power consumption
  comparison of distributed video codec and {H.264/AVC} in low-complexity
  mode,'' in \emph{18th SoftCOM}, 2010, pp. 66--70.

\bibitem{pereira2009chapter}
F.~Pereira, C.~Brites, and J.~Ascenso, \emph{CHAPTER 8 - Distributed Video
  Coding: Basics, Codecs, and Performance}, P.~L. Dragotti and M.~Gastpar,
  Eds.\hskip 1em plus 0.5em minus 0.4em\relax Boston: Academic Press, 2009.

\bibitem{artigas2007discover}
X.~Artigas, J.~Ascenso, M.~Dalai, S.~Klomp, D.~Kubasov, and M.~Ouaret, ``The
  {DISCOVER} codec: Architecture, techniques and evaluation,'' in \emph{PCS},
  2007.

\bibitem{varodayan2006rate}
D.~Varodayan, A.~Aaron, and B.~Girod, ``Rate-adaptive codes for distributed
  source coding,'' \emph{Signal Processing}, vol.~86, no.~11, pp. 3123--3130,
  2006, special Section: Distributed Source Coding.

\bibitem{zhou2019distributed}
J.~Zhou, Y.~Fu, Y.~Yang, and A.~T. Ho, ``Distributed video coding using
  interval overlapped arithmetic coding,'' \emph{Signal Processing: Image
  Communication}, vol.~76, pp. 118--124, 2019.

\bibitem{fang2023qary}
Y.~Fang, ``{$Q$}-ary distributed arithmetic coding for uniform {$Q$}-ary
  sources,'' \emph{IEEE Transactions on Information Theory}, vol.~69, no.~1,
  pp. 47--74, Jan. 2023.

\bibitem{arikan2009channel}
E.~Ar{\i}kan, ``Channel polarization: A method for constructing
  capacity-achieving codes for symmetric binary-input memoryless channels,''
  \emph{IEEE Trans. Inf. Theory}, vol.~55, no.~7, pp. 3051--3073, July 2009.

\bibitem{korada2010polar}
S.~B. Korada and R.~L. Urbanke, ``Polar codes are optimal for lossy source
  coding,'' \emph{IEEE Trans. Inf. Th.}, vol.~56, no.~4, pp. 1751--1768, Apr.
  2010.

\bibitem{lv2013novel}
X.~Lv, R.~Liu, and R.~Wang, ``A novel rate-adaptive distributed source coding
  scheme using polar codes,'' \emph{IEEE Commun. Lett.}, vol.~17, no.~1, pp.
  143--146, Jan. 2013.

\bibitem{yaacoub2016distributed}
C.~Yaacoub and M.~Sarkis, ``Distributed compression of correlated sources using
  systematic polar codes,'' in \emph{9th ISTC}, 2016, pp. 96--100.

\bibitem{yaacoub2017systematic}
------, ``Systematic polar codes for joint source-channel coding in wireless
  sensor networks and the internet of things,'' \emph{Procedia Computer
  Science}, vol. 110, pp. 266--273, 2017.

\bibitem{chiu2012hybrid}
C.-C. Chiu, S.-Y. Chien, C.-H. Lee, V.~S. Somayazulu, and Y.-K. Chen, ``Hybrid
  distributed video coding with frame level coding mode selection,'' in
  \emph{19th IEEE ICIP}, 2012, pp. 1561--1564.

\bibitem{ascenso2005improving}
J.~Ascenso, C.~Brites, and F.~Pereira, ``Improving frame interpolation with
  spatial motion smoothing for pixel domain distributed video coding,'' in
  \emph{5th {EURASIP} conference on speech and image processing, multimedia
  communications and services}, Jan. 2005, pp. 1--6.

\bibitem{cheng2005successive}
S.~Cheng and Z.~Xiong, ``Successive refinement for the {Wyner-Ziv} problem and
  layered code design,'' \emph{IEEE Transactions on Signal Processing},
  vol.~53, no.~8, pp. 3269--3281, Aug. 2005.

\bibitem{brites2008correlation}
C.~Brites and F.~Pereira, ``Correlation noise modeling for efficient pixel and
  transform domain {Wyner–Ziv} video coding,'' \emph{IEEE Trans. Circuits
  Syst. Video Technol.}, vol.~18, no.~9, pp. 1177--1190, Sept. 2008.

\bibitem{wachsmann99multilevel}
U.~Wachsmann, R.~F.~H. Fischer, and J.~B. Huber, ``Multilevel codes:
  Theoretical concepts and practical design rules,'' \emph{IEEE Trans. Inf.
  Theory}, vol.~45, no.~5, pp. 1361--1391, July 1999.

\bibitem{kubasov2007optimal}
D.~Kubasov, J.~Nayak, and C.~Guillemot, ``Optimal reconstruction in {Wyner-Ziv}
  video coding with multiple side information,'' in \emph{IEEE 9th Workshop on
  Multimedia Signal Processing}, 2007, pp. 183--186.

\bibitem{martins2009refining}
R.~Martins, C.~Brites, J.~Ascenso, and F.~Pereira, ``Refining side information
  for improved transform domain {Wyner-Ziv} video coding,'' \emph{IEEE Trans.
  Circuits Syst. Video Technol.}, vol.~19, no.~9, pp. 1327--1341, Sept. 2009.

\bibitem{miloslavskaya2014sequential}
V.~Miloslavskaya and P.~Trifonov, ``Sequential decoding of polar codes,''
  \emph{IEEE Commun. Lett.}, vol.~18, no.~7, pp. 1127--1130, July 2014.

\bibitem{trifonov2022multilevel}
P.~Trifonov, ``Design of multilevel polar codes with shaping,'' in \emph{IEEE
  ISIT}, 2022, pp. 2160--2165.

\bibitem{tal2013how}
I.~Tal and A.~Vardy, ``How to construct polar codes,'' \emph{IEEE Trans. Inf.
  Theory}, vol.~59, no.~10, pp. 6562--6582, Oct. 2013.

\bibitem{mori2009performance}
R.~Mori and T.~Tanaka, ``Performance of polar codes with the construction using
  density evolution,'' \emph{IEEE Commun. Lett.}, vol.~13, no.~7, pp. 519--521,
  July 2009.

\bibitem{trifonov2012efficient}
P.~Trifonov, ``Efficient design and decoding of polar codes,'' \emph{IEEE
  Trans. Commun.}, vol.~60, no.~11, pp. 3221 -- 3227, Nov. 2012.

\bibitem{Xiph}
``{Xiph.org Video Test Media},'' \url{https://media.xiph.org/video/derf/},
  [Online; accessed 30.09.2022].

\bibitem{ldpcaref}
``{A fork of the Intel-NTU OpenDVC},''
  \url{https://github.com/KaiLangen/openDVC}, [Online; accessed 30.11.2022].

\bibitem{tal2015list}
I.~Tal and A.~Vardy, ``List decoding of polar codes,'' \emph{IEEE Trans. Inf.
  Theory}, vol.~61, no.~5, pp. 2213--2226, May 2015.

\bibitem{Bjontegaard2001}
G.~{Bj{\o}ntegaard}, ``Calculation of average {PSNR} differences between
  {RD}-curves,'' \emph{Technical Report {VCEG-M33}, {ITU-T} {SG16/Q6}}, 2001.

\end{thebibliography}
